\documentclass[reprint,superscriptaddress,amsmath,amssymb,aps,prb,longbibliography]{revtex4-2}

\usepackage{graphics,epsf}

\usepackage{graphicx}
\usepackage{dcolumn}
\usepackage{bm}

\usepackage[capitalise]{cleveref}
\usepackage{amsmath}
\usepackage{amssymb}
\usepackage{amsfonts}

\usepackage{epstopdf}
\usepackage{amsbsy}
\usepackage[LGR,T1]{fontenc}
\usepackage{lmodern}
\usepackage{textgreek}

\usepackage{bbm}
\usepackage{braket}
\usepackage{xcolor}
\usepackage{float}
\usepackage[utf8]{inputenc}
 \usepackage{booktabs}
\usepackage[normalem]{ulem}
 \useunder{\uline}{\ul}{}
\usepackage{array}

\newcommand{\K}{~\rm {K}}

\begin{document}

\title{Chiral charge density wave in 4Hb- and 1T-TaS$_2$: The Role of interlayer coupling}

\author{Roni Anna Gofman}
\thanks{These authors contributed equally}
\affiliation{Department of Physics, Technion, Haifa, 3200003, Israel}

\author{Abigail Dishi}
\thanks{These authors contributed equally}
\affiliation{Department of Physics, Technion, Haifa, 3200003, Israel}
\author{Hyeonhu Bae}
\affiliation{Department of Condensed Matter Physics, Weizmann Institute of Science, Rehovot 7610001, Israel}

\author{Yuval Nitzav}
\affiliation{Department of Physics, Technion, Haifa, 3200003, Israel}
	
\author{Ilay Mangel}
\affiliation{Department of Physics, Technion, Haifa, 3200003, Israel}
 
\author{Nitzan Ragoler}
\affiliation{Department of Physics, Technion, Haifa, 3200003, Israel}

\author{Sajilesh K.P.}
\affiliation{Department of Physics, Technion, Haifa, 3200003, Israel}

\author{Alex Louat}
\affiliation{Diamond Light Source, Harwell Science and Innovation Campus, Didcot, OX11 0DE, UK.}
\affiliation{SIS Neutron and Muon Source, STFC Rutherford Appleton Laboratory, Chilton, Oxfordshire OX11 0QX, U.K.}
\author{Matthew D. Watson}
\affiliation{Diamond Light Source, Harwell Science and Innovation Campus, Didcot, OX11 0DE, UK.}

\author{Cephise Cacho}
\affiliation{Diamond Light Source, Harwell Science and Innovation Campus, Didcot, OX11 0DE, UK.}

\author{Dmitry Marchenko}
\affiliation{Helmholtz-Zentrum Berlin für Materialien und Energie, Albert-Einstein-Strasse 15, 12489 Berlin, Germany}
\author{Andrei Varykhalov}
\affiliation{Helmholtz-Zentrum Berlin für Materialien und Energie, Albert-Einstein-Strasse 15, 12489 Berlin, Germany}

\author{Irena Feldman}
\affiliation{Department of Physics, Technion, Haifa, 3200003, Israel}

\author{Binghai Yan}
\affiliation{Department of Condensed Matter Physics, Weizmann Institute of Science, Rehovot 7610001, Israel}

\author{Amit Kanigel}
\affiliation{Department of Physics, Technion, Haifa, 3200003, Israel}

\begin{abstract}
We use micro--angle-resolved photoemission spectroscopy (micro-ARPES) to investigate chiral charge density waves (CDWs) in 4Hb-TaS$_2$ with micron-scale spatial resolution. In the 1T layers of 4Hb-TaS$_2$, we uncover coexisting left- and right-handed CDW domains and resolve four distinct spectral patterns arising from the interplay of chirality and rotational stacking. In contrast, bulk 1T-TaS$_2$ exhibits a uniform chirality. In addition, 4Hb-TaS$_2$ shows negligible out-of-plane dispersion of the 1T-derived bands, in contrast to the pronounced interlayer coupling observed in bulk 1T-TaS$_2$. Density functional theory (DFT) calculations corroborate this picture, revealing that the interlayer interaction of the chiral order in 4Hb-TaS$_2$ is nearly two orders of magnitude weaker than in the 1T polytype. Our findings establish 4Hb-TaS$_2$ as a quasi-two-dimensional platform for exploring tunable chiral CDW phenomena.
\end{abstract}

\maketitle


\section{Introduction}
\label{sec:intro}
Chirality, the property of an object being distinguishable from its mirror image, has profound implications in both fundamental physics and materials science. In crystalline solids, chirality arises when the structure breaks all mirror and inversion symmetries. This results in left- and right-handed structural variants, which are related to each by spatial inversion or reflection.

This structural chirality can lead to a variety of exotic responses, including the anomalous Hall effect \cite{nagaosa2010anomalous}, natural optical activity \cite{barron2009molecular}, and nontrivial topological phenomena \cite{kharzeev2014chiral}.

However, handedness can also arise in systems that preserve inversion symmetry, provided all mirror symmetries are broken. A more precise description of such systems is that they are pseudo-chiral \cite{PhysRevLett.129.156401,louat2024pseudochiral} or host an emergent ferro-rotational order \cite{liu2023electrical,jin2020observation}. In both cases, the system admits two distinct states that are related to one another by a mirror operation.

 This scenario is realized, for example, in certain charge-ordered states, where the spontaneous formation of a charge density wave breaks all mirror symmetries of the lattice, resulting in left- and right-handed domains \cite{louat2024pseudochiral,guo2023ferrorotational}. 
 Pseudo-chirality may result in domain-sensitive transport or strain-coupled effects \cite{yan2024structural}. Understanding such symmetry-breaking patterns is crucial for identifying new correlated phases and unraveling the full richness of emergent phenomena in quantum materials.

In the 1T polytype of tantalum disulfide (TaS$_2$), a commensurate charge density wave forms at low temperatures, characterized by a $\sqrt{13} \times \sqrt{13}$ superlattice reconstruction relative to the atomic lattice. This reconstruction involves a periodic lattice distortion in which 13 Ta atoms cluster into a "Star of David" (SoD) configuration, with one central Ta atom surrounded by two concentric rings of six atoms each. The formation of these clusters lowers the system’s symmetry by breaking the in-plane mirror operations of the underlying trigonal lattice. Notably, the CDW can adopt one of two degenerate orientations, rotated by either +13.9$^o$ or -13.9$^o$ relative to the atomic lattice vectors. These two configurations are energetically equivalent and related by mirror symmetry.

Among transition-metal dichalcogenides (TMDs), TaS$_2$ stands out for its rich polymorphism. In addition to the 1T structure discussed above, other polytypes such as 2H, 3R, 4Hb, and 6R have been extensively studied, each exhibiting a distinct set of tunable physical properties.

The 4Hb polytype of TaS$_2$ crystallizes in a centrosymmetric structure composed of alternating 1H and 1T-TaS$_2$ layers stacked along the crystallographic \( c \)-axis (See \cref{fig:fig1}a). The 1H layers adopt a trigonal prismatic coordination, and appear in two rotational variants, denoted 1H and 1H$\textquotesingle{}$, related by a 180$^\circ$ rotation about the \( c \)-axis. This rotation inverts the in-plane orientation of the trigonal prisms and restores inversion symmetry when considering the full stacking sequence. The 1T layers, which possess intrinsic inversion symmetry, are also present in two rotated forms, 1T and 1T$\textquotesingle{}$, related by a 60$^\circ$ in-plane rotation combined with a fractional translation along the \( c \)-axis. This operation corresponds to a threefold screw axis, that links adjacent 1T layers. The overall 4Hb structure thus preserves global inversion symmetry despite the local asymmetry of the 1H layers, and exhibits both translational and screw symmetries.

As monolayers or even when stacked into separate bulk crystals, the 1H and 1T phases of TaS$_2$ exhibit fundamentally different properties: the 1H layers are metallic and can host superconductivity, while the 1T layers are correlated insulators. The close proximity of these contrasting layers gives rise to a new emergent quantum state not found in either constituent alone \cite{Ribak2020chiral, Manzeli2017}.
Recent studies have shown that a significant charge transfer occurs from the 1T layers to the 1H layers, effectively doping the latter and modifying their electronic behavior \cite{Almoalem2024charge}. This proximity, where the correlated electronic state of the 1T layers influences the superconducting behavior of the adjacent 1H layers, is believed to play a key role in stabilizing exotic superconducting phases in 4Hb-TaS$_2$, including unconventional pairing mechanisms and potentially topological order \cite{Nayak2021evidence,Almoalem2022evidence, Silber2024two}.

 

In this work, we employ micro-angle-resolved photoemission spectroscopy (micro-ARPES) with a few-micron spatial resolution to investigate the electronic structure of 4Hb-TaS\(_2\). This approach enables us to distinguish between the coexisting 1H and 1T surface terminations and to directly probe the layer-specific electronic properties. By focusing on the 1T layers, we reveal the nature of the charge density wave in the 4Hb polytype and uncover key differences from the well-studied bulk 1T-TaS\(_2\).


\section{Results}
\label{sec:Experiment}

\begin{figure}[ht!]
\includegraphics[width=0.98\columnwidth]{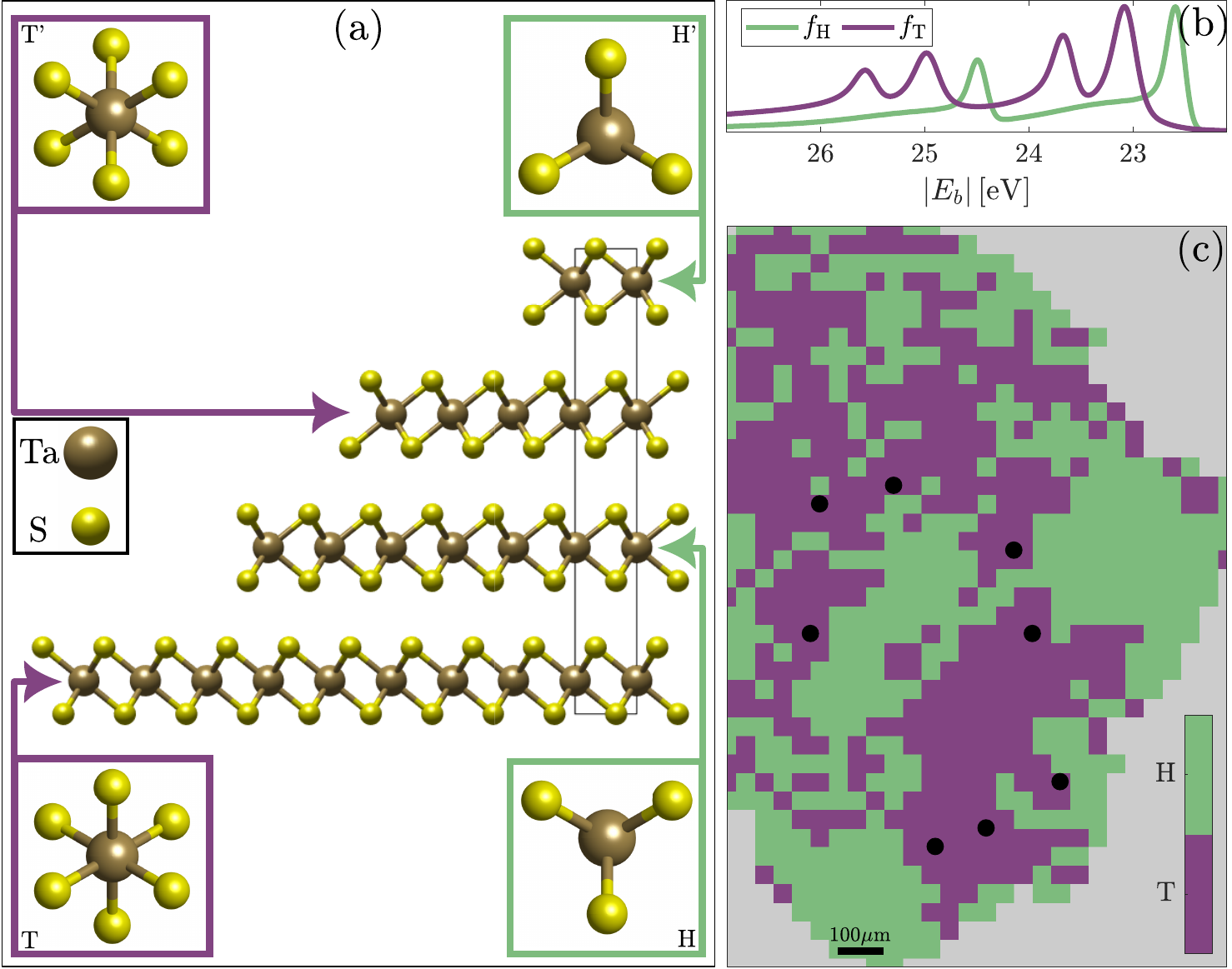}
\caption{\textbf{(a):} Crystal structure of 4Hb-TaS$_2$, including a side view and top views of each of the four distinct layer types.
\textbf{(b):}  Calculated Ta-$4f$ core-level spectra for the 2H (green) and 1T (purple) polytypes, showing their characteristic line shapes.
\textbf{(c):}  Spatial distribution map of H (green) and T (purple) domains across a $920 \times 1100\, \mu\text{m}^2$ sample area, with a pixel size of $20 \times 20\, \mu\text{m}^2$. Black circles mark positions in 1T-terminated regions where full $E(k_x, k_y)$ measurements were performed.
}
\label{fig:fig1}
\end{figure}

Upon cleaving a 4Hb-TaS\(_2\) crystal, terraces of different layers form, leading to heterogeneously terminated domains on the cleaved surface. In conventional ARPES measurements, where the spot size is typically on the order of \(100 \, \mu\text{m}\), the probed area can encompass more than one domain. As a result, the measured band structure may represent a superposition of signals from different domains, potentially including contributions from both the 1T and 1H polytypes.

To distinguish the contributions of different terminations, we employed micro-ARPES, which reduces the averaging over multiple domains inherent to conventional ARPES. With its smaller spot size, approximately \(5 \, \mu\text{m}\), this technique allowed us to scan the sample and map the electronic structure on the micrometer scale.

To establish a detailed spatial termination map of the cleaved area, we first analyzed the Ta-\(4f\) core levels using \(80 \, \text{eV}\) photons. At each point on the sample, we measured within a binding energy range of \(22 - 27 \, \text{eV}\), which contains the \(\text{Ta-}4f\) core levels. The \(\text{Ta-}4f\) core levels spectra, which is shown in Fig. \ref{fig:fig1}c,  can be used to identify the TaS\(_2\) polytype.  In the 2H polytype, which features a single Ta site, the \(\text{Ta}-4f\) spectrum displays only a spin-orbit splitting of \(1.9 \, \text{eV}\). In contrast, in the 1T phase, specifically within the C-CDW state, the \(\text{Ta-}4f\) core level line is additionally split into three components with intensity ratios of 6:6:1. This results from the distinct environments of the Ta atoms in the SoD configuration \cite{hughes1996lineshapes1,hughes1996lineshapes2,hughes1996lineshapes3}.
The spectrum of the \(\text{Ta-}4f\) core level at each point on our sample was fitted to a model line shape to determine the polytype at this location on the sample  (For more details, see SM).
In Fig. \ref{fig:fig1}b, we present a spatial map showing the distribution of different terminations where the H-terminated regions are shown in green and the T-terminated regions in purple. 

Within the T-terminated domains, we selected specific points (marked as black squares in Fig. \ref{fig:fig1}b) to perform full \(E(k_x,k_y)\) measurements, obtaining detailed band structures unique to the individual domain. This approach allowed us to analyze the properties of the 1T layers separately, minimizing cross-domain averaging.

\begin{figure*}[ht!]
\includegraphics[width=0.98\textwidth]{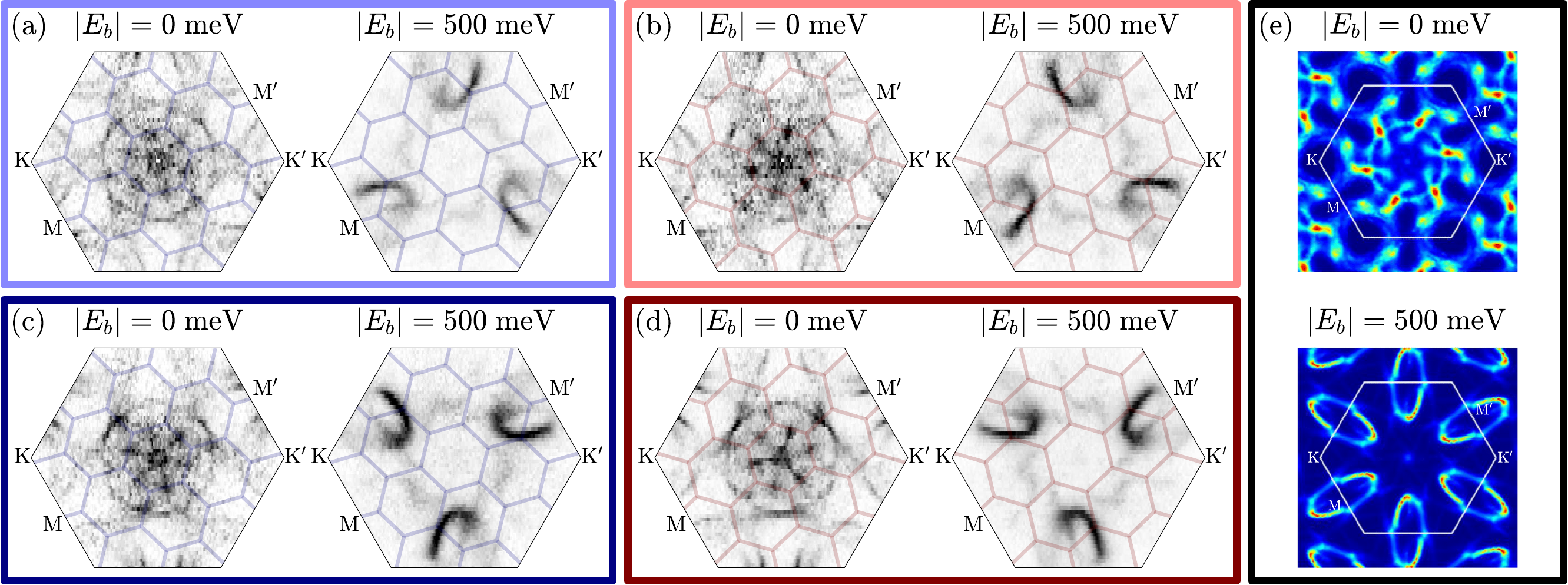}
\caption{\textbf{(a-d):} ARPES intensity maps at the Fermi level and 500 meV below the Fermi energy within the BZ are shown for the four 1T-terminated configurations. Each map has been symmetrized through a 120° rotation (see un-symmetrized data in the SM) . 
Panels within blue rectangles (a and c) represent left-handed chirality, while those within red rectangles (b and d) display right-handed chirality. The red and blue lines outline the mini-BZs associated with the CDW, rotated by -13° (blue) and +13° (red) relative to the main BZ.
Panels a and b (lighter shades) exhibit threefold symmetry, with the highest intensity centered around the M point. In contrast, Panels c and d (darker shades) are rotated by 60°, with the highest intensity centered around the M\textquotesingle{} point. \textbf{(e):} DFT calculated constant energy surfaces for a 1T termination in a 4Hb polytype at the same binding energies as in (a-d).}
\label{fig:fig2}
\end{figure*}

Surprisingly, in T-terminated domains, we observe four distinct spectral patterns. Representative data for these patterns are presented in panels (a-d) of \cref{fig:fig2}, which shows ARPES intensity maps at the Fermi level and at a binding energy (\(\left|E_b\right|=\left|E_{\rm{kin}}-E_f\right|\)) of  500\,meV.

At the Fermi level, all four patterns display a weak but clearly visible Fermi surface, reminiscent of the characteristic features of the 2H polytype. A hole pocket appears around the $\Gamma$ point, accompanied by additional hole-like features near the K and K$\textquotesingle{}$ points. Within the $\Gamma$ pocket, we detect spectral weight consistent with a Fermi surface derived from the 1T layer, suggesting partial metallization of the otherwise insulating 1T phase. This observation is consistent with previous studies reporting significant charge transfer from the 1T to the 1H layers in 4Hb-TaS$_2$~\cite{Almoalem2024charge,Nayak2021evidence}.

The 1T-derived Fermi surface includes a small pocket centered at $\Gamma$, along with six additional elongated features aligned along directions connecting the $\Gamma$-point to the centers of the surrounding CDW-induced mini Brillouin zones (BZs).  The resulting Fermi surface lacks mirror symmetry and thus exhibits chirality. We classify each domain as left- or right-handed based on the handedness of the Fermi surface, which is determined by the orientation of the CDW modulation.

The chirality is set during the formation of the SoD clusters, which define a new lattice rotated by either $+13.9^\circ$ or $-13.9^\circ$ relative to the atomic lattice vectors. To illustrate this, we overlay the corresponding CDW-induced mini BZs on the intensity maps in \cref{fig:fig2}, using red lines for the $+13.9^\circ$ configuration and blue lines for the $-13.9^\circ$ case. As indicated by the direction in which the Fermi pockets point, the two panels on the right (enclosed by light and dark red frames) exhibit right-handed chirality, while the two on the left (enclosed by light and dark blue frames) exhibit left-handed chirality.

As the binding energy increases, the spectral features associated with the 2H polytype gradually diminish, and the constant-energy maps begin to resemble those of the 1T insulating phase. Notably, the lower-energy features retain the same chirality as observed at the Fermi surface.

The differences between the four patterns are particularly evident in the intensity maps at 500\,meV binding energy. These constant-energy ARPES maps reveal an asymmetric, boomerang-like feature in momentum space, characterized by a skewed bifurcation with unequal arm lengths and intensities, further indicating broken mirror symmetry. In \cref{fig:fig2}a and c (enclosed by light and dark blue frames), the left arm is more intense, while in b and d (enclosed by light and dark red frames), the right arm dominates. This asymmetry is consistent with the handedness inferred from the orientation of the Fermi pockets.

Although only two chiralities, left- and right-handed, are possible, we observe four distinct spectral patterns in the 1T-terminated domains. This apparent doubling arises from the combination of chirality with a 60$^\circ$ rotation of the spectral intensity pattern. In the constant-energy maps at high binding energy, this manifests as asymmetric features centered either around the M point (Fig.~\ref{fig:fig2}a and b, enclosed by light blue and red) or, rotated by 60$^\circ$, around the M\textquotesingle{} point (Fig.~\ref{fig:fig2}c and d, enclosed by dark blue and red). 

We compare the experimental constant-energy maps with those calculated using Density Functional Theory (DFT) for the left-handed CDW configuration. As shown in \cref{fig:fig2}e, the DFT results closely reproduce key features of the experimental data, including the chiral structure of the 1T-derived Fermi surface within the H-associated $\Gamma$ pocket and the persistence of this chirality at lower binding energies. However, a notable discrepancy arises: while the experimental maps exhibit a threefold rotational symmetry, the DFT-calculated maps retain the full sixfold symmetry of the underlying lattice. This difference likely originates from matrix element effects in the ARPES measurements.

ARPES formally measured the single-particle spectral function times the transition dipole matrix element connecting the initial and final states of the emitted electron \cite{damascelli2004probing}. The matrix element induced modulation in the AREPS intensity contains important information about the symmetry and orbital content of the initial state. 

In a recent paper \cite{beaulieu2021unveiling}, it was shown that when using linear polarized light, a matrix element effect results in an apparent reduction of the 6-fold symmetry of the ARPES intensity to a 3-fold symmetric map in 1T-TiTe$_2$, a compound with the same structure as 1T-TaS$_2$. This is manifested by
the strongly anisotropic photoemission from Ti-3d electron
pockets.  The photoemission intensity is strong at the M\textquotesingle{} pockets but nearly vanishes at the adjacent M pockets.

Since the 4Hb polytype of TaS$_2$ belongs to the \(P6_3/mmc\) space group, characterized by six-fold rotational symmetry around the screw \(c\)-axis, adjacent pairs of 1T layers also exhibit a 60$^\circ$ rotation between them (i.e., between 1T and 1T\textquotesingle{}), as illustrated in \cref{fig:fig1}a. 
Since our measurements also used linearly polarized light, the two possible orientations of the three bright pockets can be attributed to 1T- and 1T\textquotesingle{}-terminated domains.
As shown in \cref{fig:fig2}, we observe both left- and right-chirality spectral patterns in the 1T and 1T\textquotesingle{} domains,  indicating that there is no simple correlation between the layer type (1T or 1T\textquotesingle{}) and the CDW angle.  

\begin{figure}[ht!]
\includegraphics[width=0.76\columnwidth]{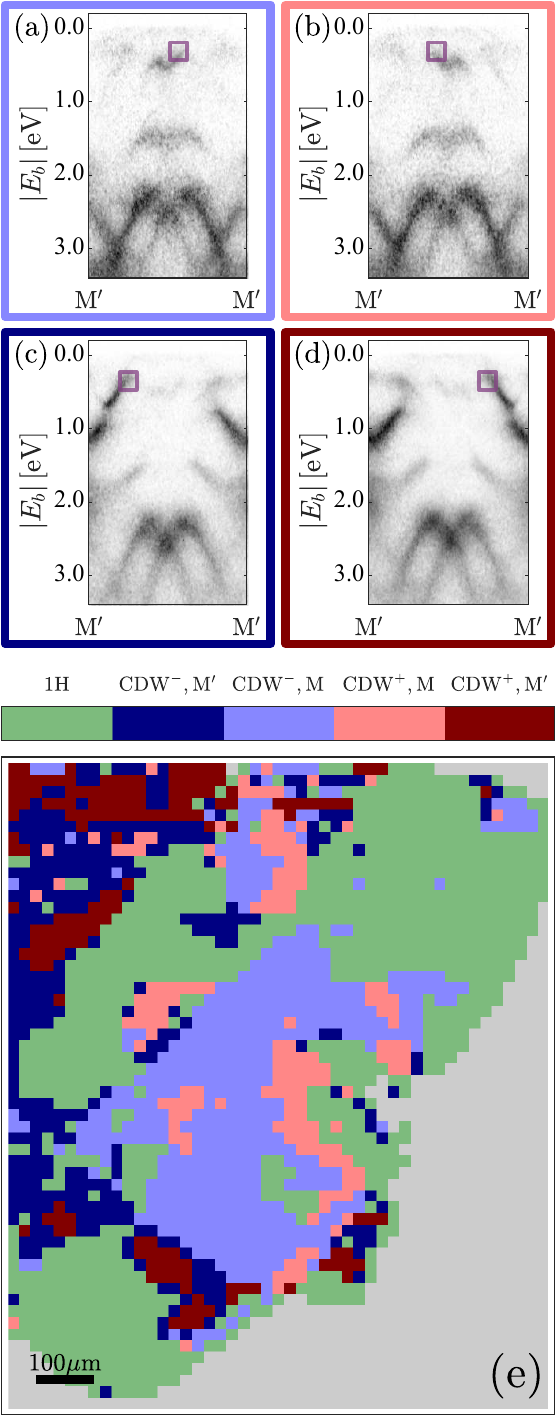}
\caption{  \textbf{(a-b):} ARPES spectra acquired along the low-symmetry \(k\)-space direction M\textquotesingle{}-M\textquotesingle{} (clockwise), for the four distinct 1T-terminated surface configurations. The colors of the frames correspond to those used in \cref{fig:fig2}.
\textbf{(e):} Part of the spatial map in \cref{fig:fig1}b (pixel size $5 \times 5\, \mu\text{m}^2$) now showing the distribution of the four configurations, with red and blue shades matching those of the frames in (a) to  (d) and in \cref{fig:fig2}. The green regions are H-terminated, while the grey areas are outside the cleaved sample. This spatial map was scanned with the detector aligned along the low symmetry line M\textquotesingle{}-M\textquotesingle{}, and intensity differences between the purple rectangles identified each 1T-termination form. }
\label{fig:fig3}
\end{figure}

To assess the prevalence of each 1T spectral pattern, we use micro-ARPES to map the sample. While it is difficult to identify the handedness from the spectra along a high symmetry line (\cref{fig:HighSymLines}),  the low-symmetry cut from \( \text{M\textquotesingle{}} \) to \( \text{M\textquotesingle{}} \) allows a clear determination of the chirality and the relative rotation ( T vs. T\textquotesingle{}). So, at each point, we measured the dispersion in momentum space along this direction.  In \cref{fig:fig3}(a-d), we present the corresponding spectra for each of the four 1T spectral patterns, with colors matching those in \cref{fig:fig2}.  

In \cref{fig:fig3}e, we show a map illustrating the distribution of 1T patterns across the sample. In this region, approximately \( 71\% \) of the 1T domains exhibit left-handed chirality (dark and light blue), and around \( 57\% \) display a 3-fold symmetry centered at the M-point (light blue and red). In another sample, these values were \( 78\% \) and \( 55\% \), respectively.  

For other samples, we did not perform a full spatial scan at this orientation but instead selected several points within the 1T domains and identified the spectral patterns at each location. Across all our samples, we consistently observed all four spectral forms: left- and right-handed CDW states in both 1T and 1T\textquotesingle{} domains.

\begin{figure}[ht!]
\includegraphics[width=0.88\columnwidth]{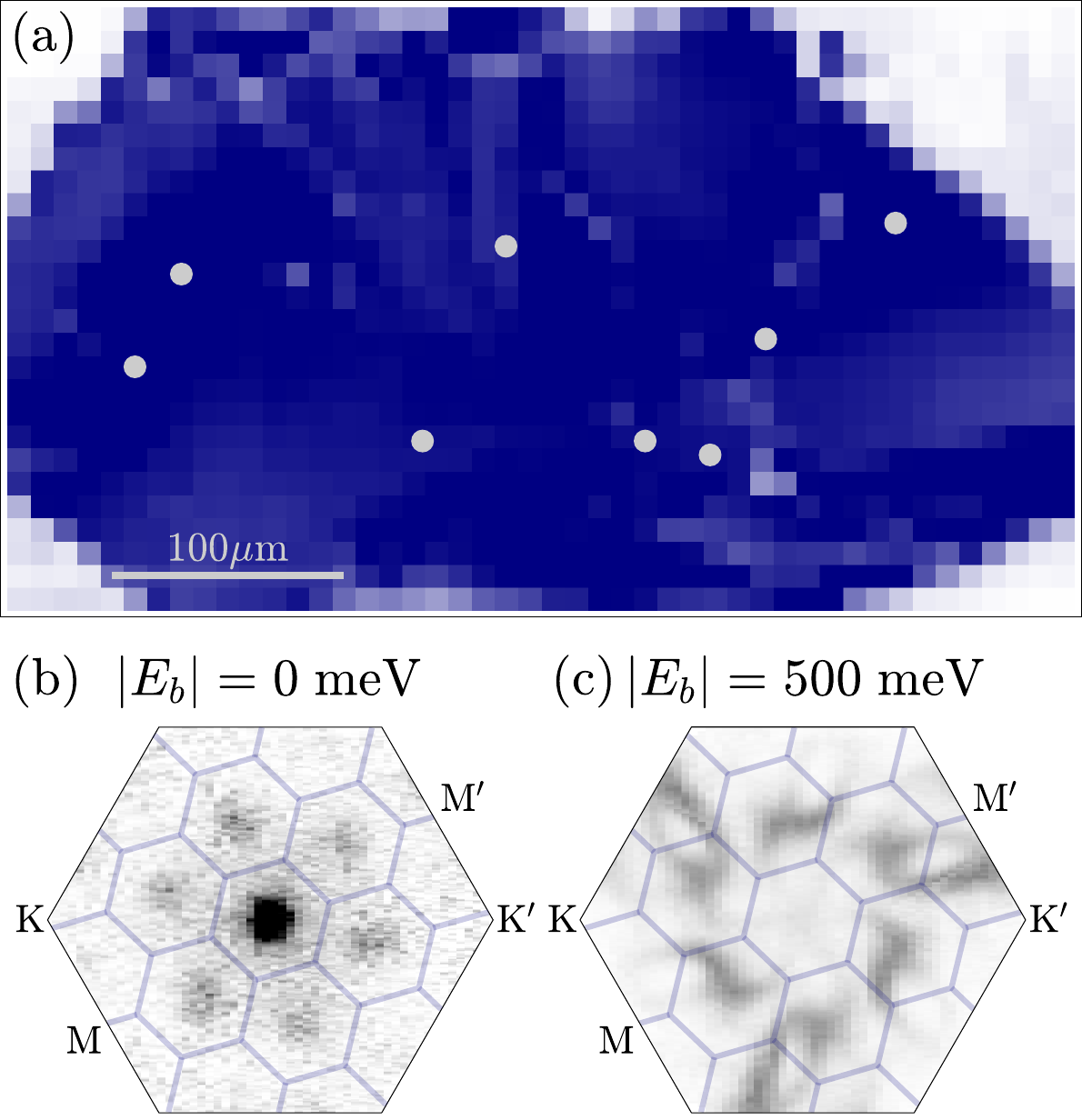}
\caption{\textbf{(a):} A spatial map of a bulk 1T-TaS$_2$ sample is. The color represents the total intensity summed over the ARPES detector. The blue region corresponds to a cleaved area of approximately \(250 \times 450 \, \mu m^2\) with a pixel size of \(10 \times 10 \, \mu m^2\). The circles indicate points where a full \(E(k_x, k_y)\) scan was measured. \textbf{(b and c):} ARPES intensity maps, symmetrized with a \(120^\circ\) rotation at \(E_b = 0\) and \(500 \, \text{meV}\), respectively.  All eight intensity maps measured at different location on the sample  exhibit the spectral pattern.}
\label{fig:fig4}
\end{figure}

The behavior we found in the 1T layers of 4Hb-TaS$_2$ should be compared to the case of bulk 1T-TaS$_2$.  
In \cref{fig:fig4}(a), we show a total-intensity map measured using micro-ARPES.  The circles represent points at which we performed a full \(E(k_x, k_y)\) measurement and determined the corresponding spectral patterns.
In \cref{fig:fig4}(b,c), we present representative constant-energy intensity maps at the Fermi level (b) and at a binding energy of \(500 \, \text{meV}\) (c). In this sample, all measurements consistently reveal a mono-domain structure exhibiting left-handed chirality and a three-fold intensity symmetry centered at the M\textquotesingle{} point. Remarkably, all 1T-TaS$_2$ samples we examined display uniform chirality across the entire measured area, which can span several mm$^2$. Raman measurements support this, confirming homogeneous chirality over large area in 1T-TaS$_2$ \cite{liu2023electrical,PhysRevLett.129.156401,Yang2022}. This stands in sharp contrast to the multi-domain chiral structure observed in all 4Hb-TaS$_2$ samples.

\begin{figure}[ht!]
\includegraphics[width=0.98\columnwidth]{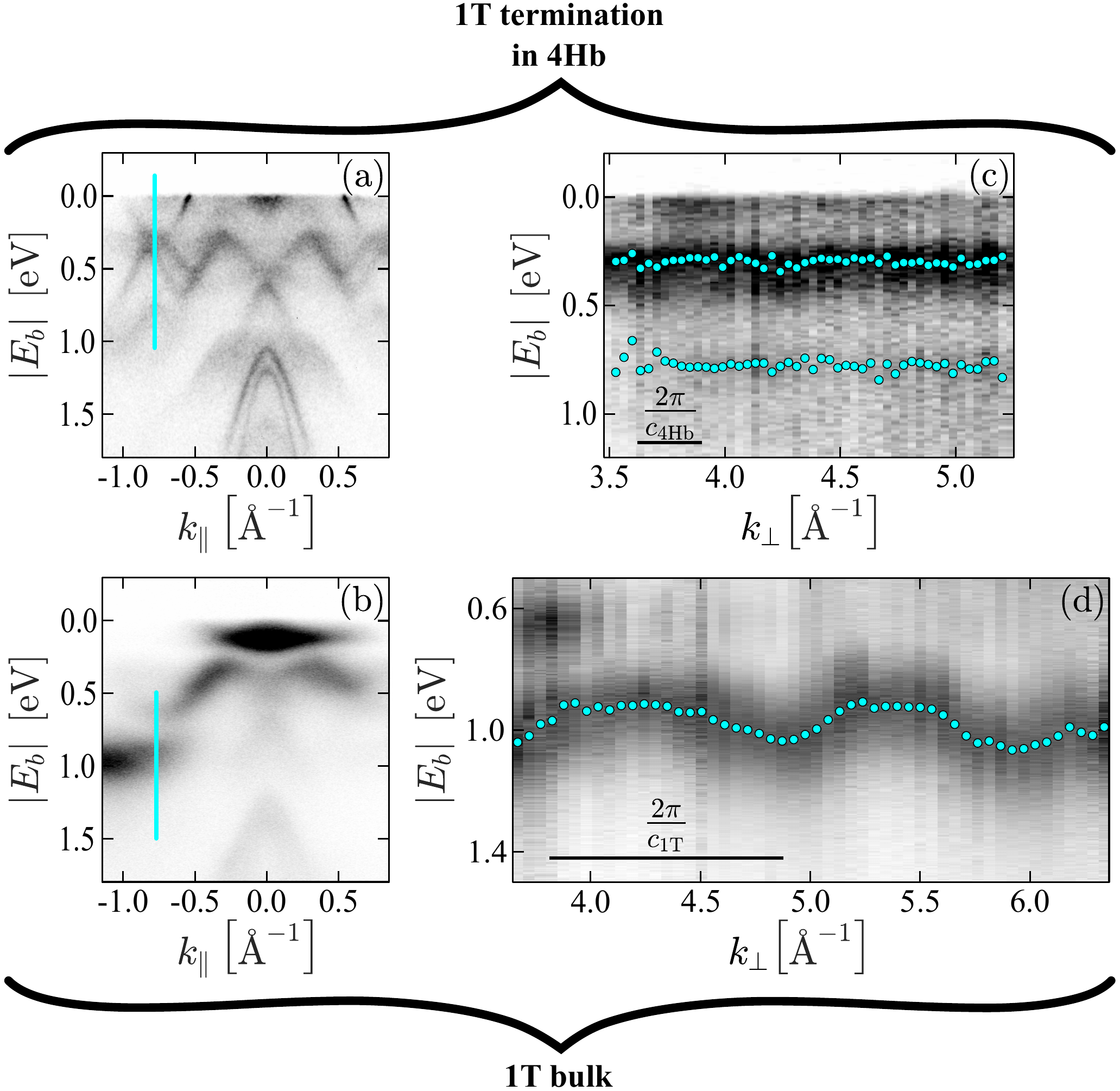}
\caption{\textbf{(a)} Dispersion along the surface-projected K$^{\prime}$–$\Gamma$–K direction for the 1T termination in bulk 4Hb, using 62 eV photon energy. The vertical line marks the $k_{\parallel}$ value used in (c).  
\textbf{(b)} Same as (a), but for bulk 1T. The vertical line marks the $k_{\parallel}$ value used in (d).  
\textbf{(c)} $k_{\perp}$ dispersion of the 1T termination in bulk 4Hb at $k_{\parallel} = -0.77 \: \mathrm{\AA^{-1}}$, corresponding to the vertical line in (a). Cyan markers denote EDC peak positions as a function of $k_{\perp}$.  
\textbf{(d)} Same as (c), but for bulk 1T at the $k_{\parallel}$ value marked in (b).}
\label{fig:fig5}
\end{figure}

We continue our comparison of the electronic spectra between the 1T-dominated domains within the 4Hb polytype and the bulk 1T polytype, shifting our attention to the out-of-plane dispersion \( (k_\perp) \).  Micro-ARPES was employed for the measurement in the 1T-dominated domains in 4Hb-TaS$2$, while large-spot ARPES was used to measure the dispersion in bulk 1T-TaS$_2$.

In both experiments, the analyzer slit was aligned along the surface projected K\textquotesingle{}–\(\Gamma\)–K crystallographic direction. Panels (a) and (b) of \cref{fig:fig5} show the corresponding spectra for 4Hb- and 1T-TaS$_2$, respectively. To probe the \( k_\perp \) dispersion, we varied the photon energy: from 40 eV to 100 eV for 4Hb polytype, covering approximately six BZs, and from 42 eV to 144 eV for 1T, covering about 2.5 BZs. Additional details on the determination of \( k_\perp \) values are provided in the SM.

We compare the \(k_\bot\) dispersion at an in-plane momentum of \( k_{\parallel} = -0.77 \, \text{\AA}^{-1} \), moving away from the \(\Gamma\) point toward the K point, as indicated by the vertical lines in panels (a) and (b) of \cref{fig:fig5}. At this momentum, a 1T-derived band is observed at a binding energy of approximately \(1 \, \text{eV}\) in 1T-TaS\(_2\), and around \(800 \, \text{meV}\) in the 1T domain of 4Hb-TaS\(_2\). In panels  \cref{fig:fig5}(c-d), we show the corresponding \(k_\bot\) dispersion of this band in the 1T-dominated domain and the bulk 1T phase, respectively.
The markers indicate the peak energy of the energy distribution curves (EDCs) at each \(k_\bot\). In the bulk 1T-TaS\(_2\) sample, this band exhibits a pronounced out-of-plane dispersion with a bandwidth of approximately \(120 \, \text{meV}\), consistent with previous reports \cite{ngankeu2017quasi,Nitzav2024emergence} and DFT calculations \cite{Ritschel2018} . In contrast, the same band in the 1T-terminated domain of 4Hb-TaS\(_2\) shows negligible \(k_\bot\) dependence, with a bandwidth smaller than \(20 \, \text{meV}\).
We examined additional 1T-derived bands at various in-plane momenta and found that all exhibit similarly flat \(k_\bot\) dispersion in 4Hb-TaS\(_2\) (see SM).

\section{Discussion} 
The absence of out-of-plane dispersion in the 1T-dominated domains of 4Hb-TaS$_2$ indicates that the 1T layers are effectively decoupled. As we show, this has a profound impact on the morphology of the CDW.  
Comparing the electronic structure of bulk 1T-TaS$_2$ with that of the 1T domains in 4Hb thus provides a unique opportunity to contrast a fully three-dimensional CDW with a quasi-two-dimensional variant that shares the same fundamental CDW motif.

It is known that in bulk 1T-TaS$2$ the stacking of the SoD determines the transport properties of the material \cite{Ritschel2018}. The out-of-plane bandwidth in bulk 1T is larger than the in-plane bandwidth; this results in an unusual band structure for a TMD.
The interlayer coherence imposes a constraint on domain wall formation: any chiral domain wall must extend through multiple layers, making its formation both energetically and structurally less favorable \cite{zhao2023spectroscopic}.
 
Coexisting chiral domains of opposite handedness in bulk crystals could be created using light pulses, thermal quenching, or by applying voltage in thin-flake devices \cite{zong2018ultrafast,de2024dynamic,riffle2024cooling,liu2023electrical}. This could be related to stacking faults that weaken the interlayer coupling. 
  

By contrast, the 4Hb polytype exhibits much weaker interlayer coupling, effectively confining electrons within individual layers. This reduced coupling allows domain walls to form independently in each layer, without the need for alignment between them. As a result, chiral textures can emerge in a more flexible and layer-resolved manner, consistent with the observed localized domains in 4Hb. A previous study demonstrated that while directly stacked 1T layers tend to adopt identical chirality, the insertion of 1H layers can decouple adjacent 1T layers and allow them to host different chiralities \cite{husremovic2023encoding}, consistent with our findings.

Our DFT calculations further support this picture. We modeled a four-layer 4Hb-TaS$_2$ system and compared two stacking configurations: one in which both the 1T and 1T$^\prime$ layers exhibit left-handed chirality, and another in which the layers adopt opposite chiralities (For more details, see SM). The homochiral configuration is energetically favored by \( 4 \, \text{meV} \) per SoD. This relatively small energy difference suggests a weak interlayer coupling of the chiral order, allowing for the possibility of variations in chirality across layers. In contrast, in bulk 1T-TaS$_2$, the energy cost of a chirality mismatch is about \( 500 \, \text{meV} \) per SoD \cite{zhao2023spectroscopic}, which imposes a rigid interlayer alignment and precludes the formation of independent, layer-confined chiral domain walls. 

All examined 4Hb samples exhibit the four characteristic 1T spectral patterns, though in varying proportions. Using ARPES, we cannot distinguish between a scenario in which each 1T layer adopts a uniform chirality and the observed domain walls result from surface terraces, and one in which oppositely chiral domains coexist within the same 1T layer. However, a recent STM study on the same samples directly observed a domain wall separating left- and right-handed chiralities within a single 1T layer \cite{Silber2024two}. Our results are consistent with recent XRD measurements that revealed the simultaneous presence of both left- and right-handed \(\sqrt{13} \times \sqrt{13}\) CDW domains in 4Hb-TaS\(_2\) \cite{PhysRevB.111.L041101}.

\section{Summary}
\label{sec:summary}
We investigate chiral domains formation in the 4Hb polytype of TaS$_2$ using micro-ARPES. By resolving the contributions of the 1T and 1H layers, we examine how interlayer decoupling influences electronic behavior.  

In 1T-terminated regions, four distinct chiral CDW domains emerge, exhibiting a threefold symmetry in two orientations, corresponding to left- or right-handed chirality. This chirality arises from SoD lattice distortions in the 1T phase, which break mirror symmetry and introduce spatially varying electronic structures. While bulk 1T-TaS$_2$ typically supports a single chiral domain, 4Hb-TaS$_2$ hosts multiple chiral domains, suggesting that weak interlayer coupling allows independent chiral orientations to form in each layer. These findings highlight the role of dimensionality and interlayer interactions in controlling electronic symmetry-breaking in TMDs.  

To further explore the origin of these domains, we perform DFT calculations, which indicate that the observed multi-chiral domains likely arise from intrinsic intra-layer variations rather than independent layer-wise chirality. Additionally, k$_\perp$-dependent ARPES measurements reveal a stark contrast between the strong out-of-plane dispersion in bulk 1T-TaS$_2$ and the suppressed dispersion in 4Hb-TaS$_2$, further confirming that electrons remain largely confined within individual layers in the 4Hb polytype. This reduced interlayer coupling facilitates the formation of localized chiral domains without the constraints imposed by adjacent layers.  

Our results provide new insights into the interplay between charge order, symmetry breaking, and electronic correlations in layered quantum materials. The presence of multi-chiral domains in 4Hb-TaS$_2$ offers a unique platform to explore unconventional superconductivity, topological phases, and emergent quantum states in TMDs.

\section{Methods}
\label{sec:ethod}
4Hb crystals were synthesized using the chemical vapor transport technique. Se was incorporated during growth to stabilize the 4Hb phase and enhance the properties of the superconducting phase. Se concentrations in the analyzed samples ranged from 0\% to 5\%. The grown crystals were then cut into pieces, typically a few millimeters wide and approximately 0.5 mm thick.

Micro-ARPES measurements were performed at the I05 beamline of the Diamond Light Source. Samples were introduced into an ultra-high vacuum (UHV) chamber and maintained at approximately \(40\K\). They were cleaved \emph{in situ} to ensure clean surfaces. The beam, with a spot size of about \(5\times5\;\mu\mathrm{m}^2\) and linear horizontal (LH) polarization, was used with photon energies ranging from 60 to 100 eV.

Out-of-plane ARPES measurements on bulk 1T samples were conducted at the ARPES One-Square beamline at the BESSY-II synchrotron. These experiments were carried out under UHV conditions, with pressures maintained below \(5\times 10^{-11}\) torr, and at an approximate temperature of \(40\K\). Beam spot size was about 100 microns.  

Density functional theory (DFT) calculations were performed with the \textsc{VASP} software using the projector-augmented wave method \cite{Kresse1999PRB}. We employed the Perdew-Burke-Ernzerhof (PBE) exchange-correlation functional \cite{Perdew1996PRL}, while van der Waals interactions were accounted for via the DFT-D3 scheme with Becke-Johnson damping \cite{Grimme2011JCC}. To model the 4Hb stacking, we constructed $\sqrt{13} \times \sqrt{13}$ supercells for bulk 4Hb-TaS$_2$ and for 1T/1H bilayer structures, choosing either left- or right-handed chirality. For the heterochiral 4Hb-TaS$_2$ calculation, see SM. The bilayer model was used to compute the Fermi surface of the exposed 1T layer. Spin-orbit coupling was included in these Fermi surface calculations.
A plane wave energy cutoff of 400~eV was adopted. The electronic energy convergence criterion was set to $10^{-7}$~eV. Ionic positions were relaxed until all residual Hellmann-Feynman forces were smaller than 1~meV/\AA. Band structure unfolding was carried out using the \textsc{VASPKIT} code \cite{Wang2021CPC}.


\section{Acknowledgments}
Part of this work was carried out with the support of Diamond Light Source, instrument i05 (proposal SI33131).
We thank Helmholtz-Zentrum Berlin\textquotesingle{}s ARPES One-Square beamline for allocating synchrotron radiation beamtime. The work at the Technion was supported by the Israeli Science Foundation under grant number ISF-1263/21.

\bibliographystyle{apsrev4-2}
\bibliography{MainBib}

\clearpage

\appendix
\onecolumngrid
\crefname{section}{Supplementary Material}{Supplementary Materials}
\section*{Supplementary Material}
\addcontentsline{toc}{section}{Supplementary Material}
\refstepcounter{section}
\label{supp}

\renewcommand{\thesubsection}{Sup. \Alph{subsection}}

\subsection{H and T domains}
\label{Appendix:CoreLevelFit}
\setcounter{figure}{0} 
\renewcommand{\thefigure}{\thesubsection.\arabic{figure}}

Following the methods and parameters described in \cite{hughes1996lineshapes1}, \cite{hughes1996lineshapes2}, and \cite{hughes1996lineshapes3}, we constructed the Ta-\(4f\) core-level line shapes for the 1T and 1H polytypes in 4Hb-TaS$_2$.  Each line shape accounts for a spin-splitting of 1.9 eV and maintains an amplitude ratio of 2:1 between the \(4f_{7/2}\) and \(4f_{5/2}\) components. Additionally, the 1T line shape incorporates CDW splitting, where the peaks from atoms in positions "b" and "c" are enhanced by a factor of 6, making the peak from atoms in position "a" negligible.

The total spectrum of the Ta \(4f\) core level in 4Hb comprises contributions from both polytypes:  

\[
    f_{4\rm{Hb}} (E_b) = a_{\rm{H}} f_{\rm{H}} (E_b) + a_{\rm{T}} f_{\rm{T}} (E_b)
\]

where \(a_{\rm{H}}\) and \(a_{\rm{T}}\) represent the amplitudes of the H and T components, respectively. These amplitudes vary depending on the local composition of H and T domains, photon flux, and measurement duration.  

We fitted the measured spectrum to \(f_{4\rm{Hb}} (E_b)\), incorporating an additional parabolic background. The relative percentages of H and T components were determined using the following equations:  

\[
\begin{array}{c@{\quad\quad\quad\quad}c}
    \%H = \frac{a_{\rm{H}}}{a_{\rm{H}} + a_{\rm{T}}} \times 100\% &
    \%T = \frac{a_{\rm{T}}}{a_{\rm{H}} + a_{\rm{T}}} \times 100\%
\end{array}
\]

Regions with a higher percentage of the H component were identified as H-domains, while regions with a higher percentage of the T component were classified as T-domains. 

\subsection{Un-symmetrized Data}
\label{Appendix:OriginalData}
\setcounter{figure}{0} 
\renewcommand{\thefigure}{\thesubsection.\arabic{figure}}
\begin{figure*}[ht!]
\includegraphics[width=0.98\textwidth]{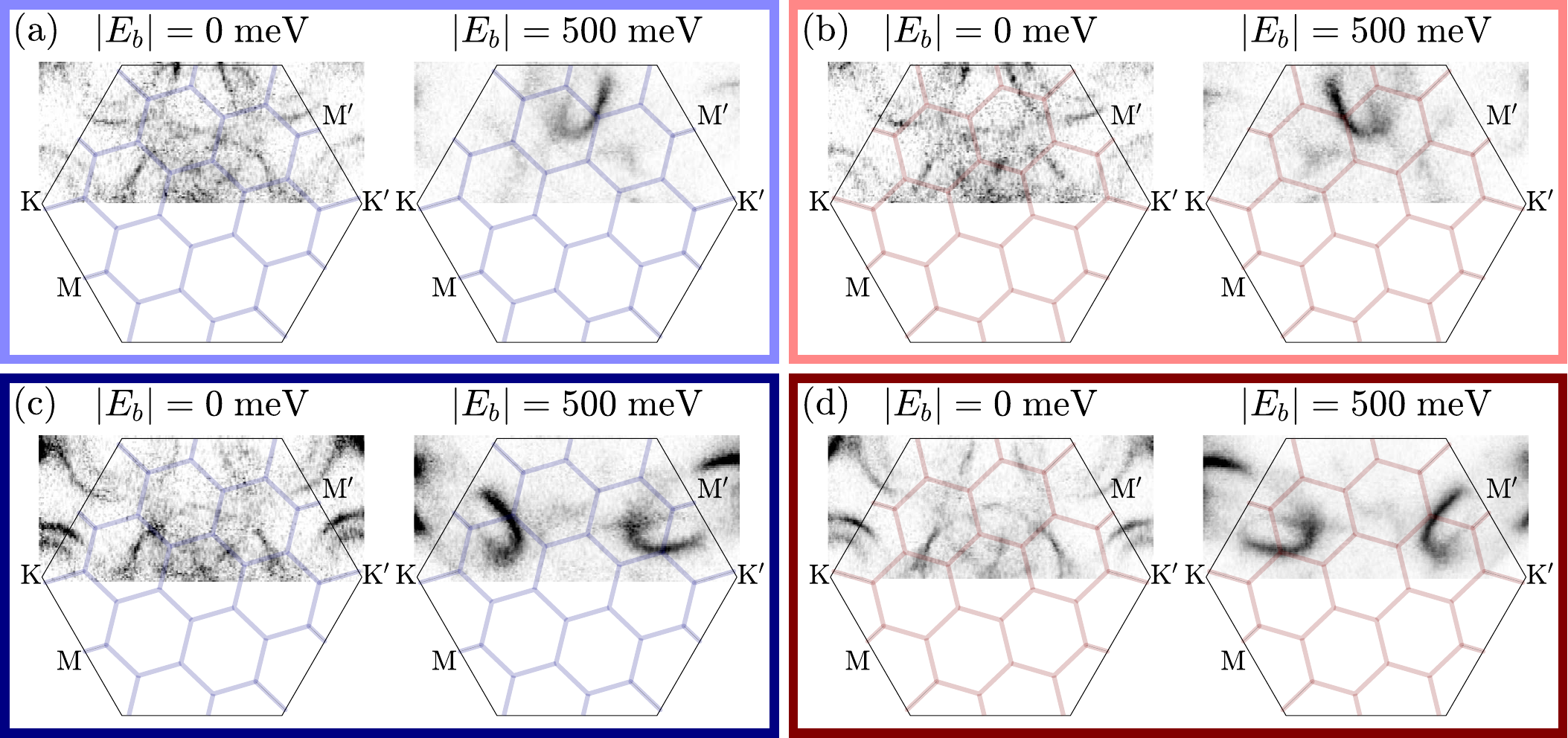}
\caption{ \textbf{(a–d):} \textit{As measured}, energy cuts at the Fermi level and 500\,meV below the Fermi level within the Brillouin zone (BZ) for the four 1T-terminated configurations.  
Panels enclosed in blue (a and c) correspond to left-handed chirality, while those in red (b and d) correspond to right-handed chirality. The red and blue lines indicate the mini-BZs associated with the charge density wave (CDW), rotated by +13\(^\circ\) (red) and –13\(^\circ\) (blue) relative to the main BZ.  
Panels a and b (lighter shading) exhibit threefold symmetry with intensity maxima near the M point, whereas panels c and d (darker shading) are rotated by 60\(^\circ\), with intensity centered around the M\textquotesingle{} point.
}
\label{fig:Original}
\end{figure*}

\newpage
\subsection{High Symmetry Lines}
\label{Appendix:HighSymmetryLines}
\setcounter{figure}{0} 
\renewcommand{\thefigure}{\thesubsection.\arabic{figure}}
We examine the four patterns of 1T-termination ARPES spectra along all high symmetry lines.


\begin{figure}[ht!]
    \includegraphics[width=0.95\columnwidth]{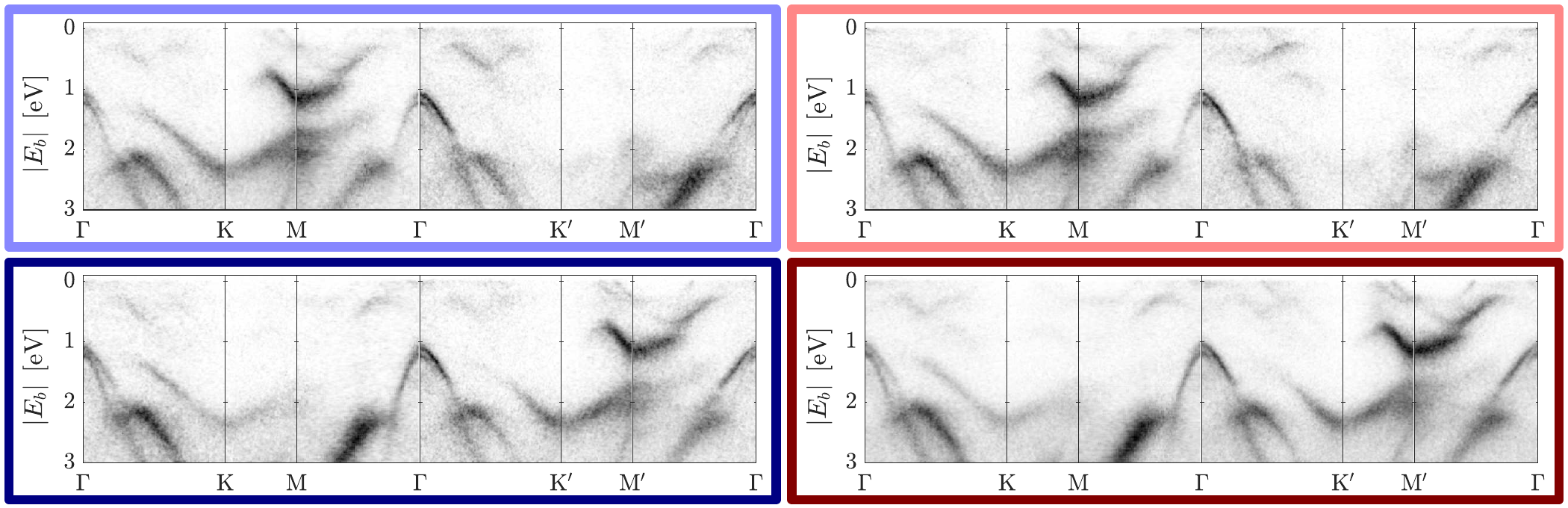}
    \caption{ARPES band structure of the 1T termination along the high-symmetry paths $\Gamma$-K, K-M, M-$\Gamma$, $\Gamma$-K\textquotesingle{}, K\textquotesingle{}-M,\textquotesingle{} and M\textquotesingle{}-$\Gamma$, shown for each of the four spectral patterns of the 1T termination. The frame colors correspond to those in \cref{fig:fig2}, where the high-symmetry points are also indicated.}
    \label{fig:HighSymLines}
\end{figure}


\newpage
\subsection{Out-of-Plane Dispersion Analysis}
\label{Appendix:KZ}
\setcounter{figure}{0} 
\renewcommand{\thefigure}{\thesubsection.\arabic{figure}}
To extract the out-of-plane dispersion (\(k_\perp\)), we assumed a free-electron final state with an inner potential \(V_0 = 9\,\mathrm{eV}\) for the 4Hb samples \cite{Almoalem2024charge} and \(V_0 = 15.4\,\mathrm{eV}\) for the 1T bulk samples \cite{Nitzav2024emergence}. The value of \(k_\perp\) is determined for each photon energy, kinetic energy, and emission angle using
\[
k_\perp = \frac{1}{\hbar} \sqrt{2m_e \left(E_k \cos^2 \vartheta + V_0 \right)},
\]

where \(E_k\) is the photoelectron kinetic energy and \(\vartheta\) is the emission angle relative to the surface normal.

\begin{figure}[ht!]
\includegraphics[width=0.7\textwidth]{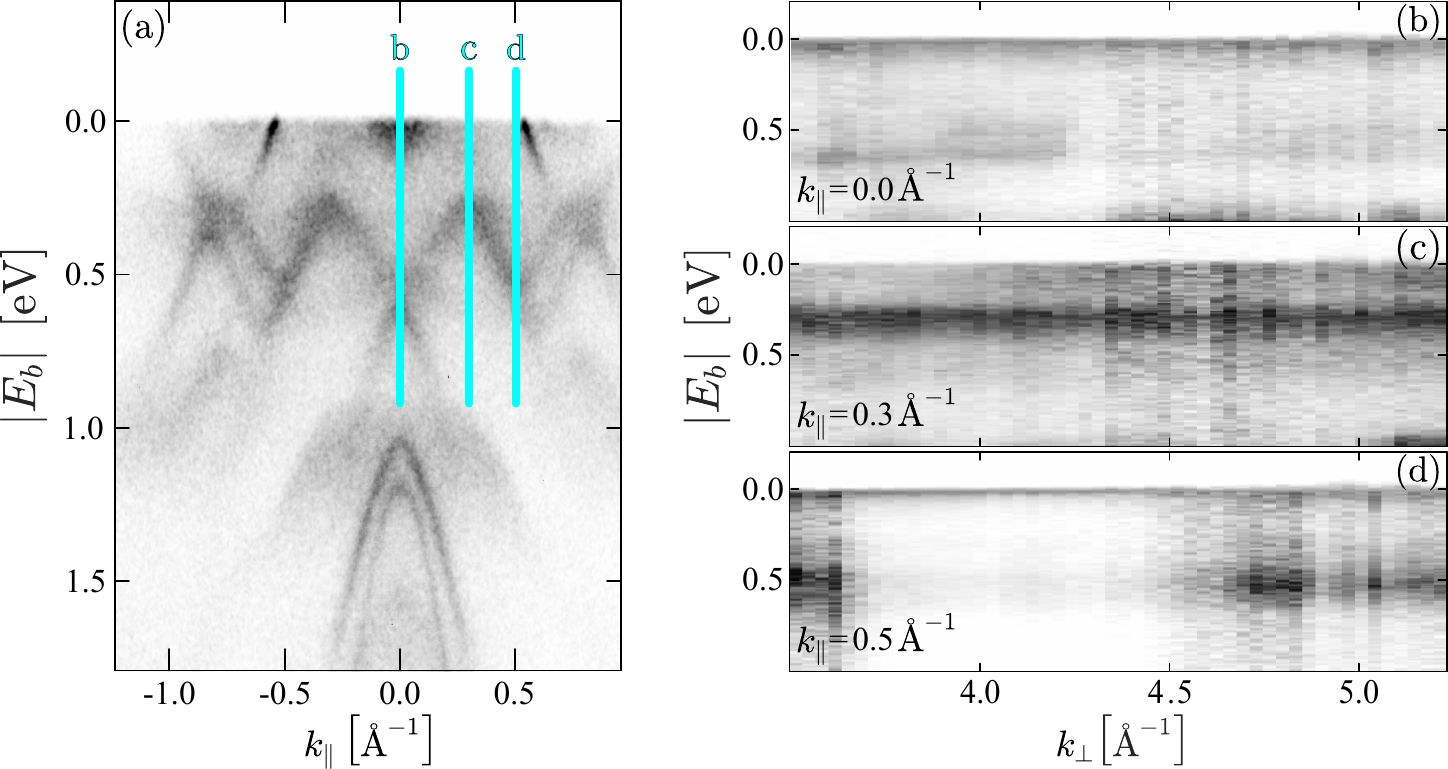}
\caption{ (a) K$^{\prime}$–$\Gamma$–K dispersion with vertical lines marking the momenta at which \(k_\perp\) dispersion was measured for representative bands of interest.  
(b–d) \(k_\perp\) dispersion measured at the \(\Gamma\)-point, \(k_\parallel = 0.3\,\text{\AA}^{-1}\), and \(k_\parallel = 0.5\,\text{\AA}^{-1}\), respectively. The bands are flat with visible intensity modulations but show no dispersion.
}
\label{fig:kzsupp}
\end{figure}
\newpage

\newpage
\subsection{Energy Difference between homochiral and heterochiral 4Hb-TaS$_2$}
\label{Appendix:DFT}
\setcounter{table}{0} 
\renewcommand{\thetable}{\thesubsection.\arabic{table}}
The heterochiral 4Hb-TaS$_2$ must be represented in a $13 \times 13$ supercell that contains two 1T layers of opposite CDW handedness, each built from the $\pm 13.9^\circ$-rotated $\sqrt{13} \times \sqrt{13}$ units~\cite{Sung2022,Husremovic2023}. The resulting Ta$_{676}$S$_{1352}$ system is computationally prohibitive for standard DFT. We therefore trained machine learning force fields (MLFFs) using \textit{ab initio} molecular dynamics (AIMD), implemented in \textsc{VASP}, on the left- and right-handed 1T/1H bilayers in a $\sqrt{13} \times \sqrt{13}$ supercell and used them to evaluate the total energies of three bulk models in $13 \times 13$ superlattices: (1) L/L (homochiral-left), (2) R/R (homochiral-right), and (3) L/R (heterochiral). 

For the AIMD simulations with PBE-D3, a total of 245 (224) snapshots for left- (right-) handed 1T/1H bilayers were extracted from 1~ps trajectories, which used a 0.1~fs integration time-step (105 steps). These trajectories were run at 100~K in the NVT ensemble, employing a Langevin thermostat with a friction coefficient of 5~ps$^{-1}$ for all atoms. The cutoff radii for twelve radial descriptors and eight angular descriptors on each atom were set to 8~\AA{} and 5~\AA{}, respectively.

The L/R 4Hb-TaS$_2$ was constructed by substituting the top 1T layer of R/R with the left-handed 1T, without further relaxation. Total energies of these three homochiral and heterochiral structures were evaluated using two force fields trained on left- and right-handed bilayers for cross-validation. The result is shown in Table~\ref{tab:energy_differences}. Two independently trained force fields predicted the cross-handed bilayer energies to agree within 1~$\mu$eV/atom, allowing us to treat the small mismatch of $-0.57$~meV/Ta$_{13}$S$_{26}$ in homochiral 4Hb structures as a systematic error when applying bilayer-trained force fields to the bulk. After correction, the heterochiral configuration is found to be 4.11~meV/Ta$_{13}$S$_{26}$ higher in energy than the homochiral ground state, corresponding to $\sim$1.2~K/atom. Hence, homochiral and heterochiral domains are expected to coexist, with the final handedness likely quenched during the near-room-temperature CDW lock-in. A layer-by-layer handedness flip requires concerted Ta displacements and is therefore unlikely at cryogenic temperatures, as in the case of stacked 1T-TaS$_2$~\cite{Lee2019Origin}.

\begin{table}[ht]
\centering
\begin{tabular}{@{}>{\centering\arraybackslash}p{6cm}%
                >{\centering\arraybackslash}p{3.2cm}%
                >{\centering\arraybackslash}p{3.2cm}%
                >{\centering\arraybackslash}p{1.6cm}@{}}
\toprule
\textbf{MLFF Training Set} & \textbf{(L)1T/1H-TaS$_2$} & \textbf{(R)1T/1H-TaS$_2$} & \textbf{Average} \\ \midrule
(L)1T/1H-TaS$_2$ vs. (R)1T/1H-TaS$_2$  & +0.03 & -0.04 & -0.01 \:\:\:\:\: \\
(L/L)4Hb-TaS$_2$ vs. (R/R)4Hb-TaS$_2$  & -0.35 & -0.79 & -0.57 (a) \\
(L/L)4Hb-TaS$_2$ vs. (L/R)4Hb-TaS$_2$  & -6.61 & -2.18 & -4.39 (b)\\ \midrule
\multicolumn{3}{l}{Energy difference after correction of systematic error: (b) - (a)/2} & -4.11 \\
\bottomrule
\end{tabular}
\caption{Total energy differences (in meV per Ta$_{13}$S$_{26}$) for bilayer 1T/1H-TaS$_2$ and bulk 4Hb-TaS$_2$ structures with different CDW chiralities. ``L'' (``R'') denotes a left- (right-) handed CDW in the 1T layer. Energies were computed using MLFFs trained separately on left- and right-handed 1T/1H bilayers. A negative energy difference $\Delta E = E_1 - E_2$ indicates that configuration 1 is thermodynamically more stable than configuration 2.}

\label{tab:energy_differences}
\end{table}

\end{document}